# Ensemble inequivalence in long-range quantum spin systems

Daniel Arrufat-Vicente [1,*] David Mukamel [2] Stefano Ruffo [3,4,5] and Nicolò Defenu [1]

[1] *Institute for Theoretical Physics, ETH Zürich, Wolfgang-Pauli-Strasse 27, 8093 Zürich, Switzerland*

[2] *Department of Physics of Complex Systems, Weizmann Institute of Science, Rehovot 7610001, Israel*

[3] *Istituto dei Sistemi Complessi, Consiglio Nazionale delle Ricerche, Via Madonna del Piano 10, I-50019 Sesto Fiorentino, Italy*

[4] *INFN Sezione di Firenze via Sansone 1, 50019 Sesto Fiorentino (FI), Italia*

[5] *SISSA, Via Bonomea 265, I-34136 Trieste, Italy*

(Received 22 April 2025; accepted 13 January 2026; published 27 March 2026)

Ensemble inequivalence, where a system's thermodynamic properties depend on the statistical ensemble used to describe it, is a well-established phenomenon in classical systems with long-range interactions. Here, we present a detailed analysis of a quantum ferromagnet spin model that exhibits this same behavior. We find that the microcanonical and canonical phase diagrams are identical at zero temperature ($T = 0$), but differ significantly at finite temperatures. Our results highlight a breakdown of ensemble equivalence in this quantum long-range model, contrasting with what is observed for short range. These findings have important implications for the study and manipulation of long-range interacting quantum platforms, such as those used in atomic, molecular, and optical physics.

## I. INTRODUCTION

The study of the equilibrium and dynamical behavior of long-range interacting quantum systems has recently attracted significant attention of the research community. This interest is partly driven by advances made in the observation, manipulation, and control of atomic, molecular, and optical (AMO) systems, where long-range interactions among microscopic constituents are a central feature [1–8]. Interactions are typically classified as long range whenever the two-body interaction potential $V(r)$ between the microscopic constituents decays as a power law of the distance $r$, $V(r) \propto r^{-\alpha}$, with $\alpha > 0$ and sufficiently small.

The behavior of such systems is deeply influenced by the exponent $\alpha$. When $\alpha > \alpha_*$, where $\alpha_*$ is a universal threshold, the critical behavior of a system in equilibrium resembles that of a short-range interacting system. Conversely, in $d$-dimensional systems with $d < \alpha < \alpha_*$, the scaling near phase transitions is altered by the long-range couplings [9–16]. Finally, when $\alpha < d$, which is the strong long-range regime, standard thermodynamics breaks down. To recover the extensivity of energy, a rescaling factor, often referred to as Kac's rescaling, becomes necessary. This, however, does not ensure additivity of thermodynamic functions, resulting in unusual thermodynamic behavior.

*Contact author: darrufat@phys.ethz.ch

In classical systems, this latter regime is characterized by hallmark phenomena such as *quasistationary states* (QSSs) [17–20] and *ensemble inequivalence* [21]. Quasistationary states are metastable states that slowly decay to the equilibrium state on timescales that diverge with system size. Ensemble inequivalence, on the other hand, refers to the fact that the thermodynamic behavior of the system becomes ensemble dependent. This feature is the central focus of the present work. Ensemble inequivalence takes place whenever the microcanonical entropy develops nonconcave regions with respect to the energy so that the mapping to the canonical free energy defined by the Legendre transform is no longer one to one. For systems with short-range interactions, the nonconcave region of the entropy progressively disappears as the system size increases, driven by phase separation. In this process, the system increases its entropy while maintaining a constant energy by allowing different phases to coexist. However, for nonadditive systems, the energy cost associated with creating a surface is significant and therefore hinders phase separation. A consequence of nonconcave entropy regions is negative specific heat, a phenomenon known to occur during stellar formation [22].

In quantum systems, the statistical mechanics of strong long-range interactions remains to a large extent underexplored, with most of its progress inspired by the study of quasiclassical regimes. For instance, theoretical evidence of quantum QSSs has emerged [23–25]. Ergodicity breaking, another intriguing feature of long-range physics, is more pronounced in the microcanonical ensemble and has been observed in both classical and quantum contexts [26–29]. A comprehensive universal picture for these multifaceted theoretical evidences has just emerged [30–32].

Coming to ensemble inequivalence, a quantum spin model has been recently demonstrated to exhibit distinct canonical and microcanonical phase diagrams [33]. The model comprises quantum spins with fully connected long-range interactions and multispin coupling. It exhibits a paramagnetic to ferromagnetic phase transition that, depending on the interaction parameters, is either first order or continuous. The manifolds of the two transition types are separated by tricritical points. While the two ensembles yield identical $T = 0$ phase diagrams, their finite-temperature phase diagrams differ significantly. In Ref. [33], the critical and tricritical manifolds have been calculated, demonstrating that the two ensembles exhibit rather distinct tricritical manifolds. The first-order transition manifolds were only drawn schematically based on some general arguments.

Here, we present an alternative method for calculating the canonical free energy and the microcanonical entropy, yielding the full phase diagrams including the first-order transitions. We also show how negative specific heat emerges near the first-order transition lines. The analysis is carried out by mapping the quantum problem onto a classical one with the caveat that one has to add an additional time direction. This is done by means of a Suzuki-Trotter decomposition that allows us to disentangle noncommuting terms of the Hamiltonian. Such approach has already been used in many other setups and is believed to be exact for long-range interacting systems [34–36].

The significance of this study is underscored by the growing interest in controlling multibody interactions in quantum many-body systems, particularly within the AMO community [37,38]. Quantum tricritical points, naturally arising in such systems [39], are of special relevance. Notably, various experimental AMO platforms [8] allow for the exploration of different statistical ensembles: They can be realized as isolated microcanonical systems, such as dipolar atomic or molecular gases [40–42], or as canonical systems coupled to a thermal bath, as seen in cold atoms within optical cavities [7]. In these settings, system sizes are finite but potentially large enough to explore the phenomenology described in this study. Moreover, cavity-mediated interactions create globally flat potentials, making them ideal for experimentally testing our predictions [25,43,44]. Recent cavity QED experiments, for example, have probed the prethermalization dynamics of long-range systems [45]. It is also important to point out that fully connected quantum Hamiltonians hold promise for optimizing classical combinatorial problems via adiabatic quantum computing [46].

## II. THE MODEL

Our findings apply to a concrete example of long-range quantum system, where the extension of the classical picture to the quantum realm can be carried out explicitly. Following Ref. [33], we introduce the Hamiltonian of a long-range quantum ferromagnetic spin-1/2 chain with four-spin interactions:

$$\mathcal{H} = -\frac{J}{N}\left(\sum_\ell \sigma_\ell^z\right)^2 - h\sum_\ell \sigma_\ell^x - \frac{K}{N^3}\left(\sum_\ell \sigma_\ell^z\right)^4, \tag{1}$$

where the summations run over all sites of the lattice $\ell \in \{1, \dots, N\}$. The operators $\sigma_\ell^\mu$ correspond to the $\mu = x, y, z$ Pauli matrices at site $\ell$:

$$\sigma_\ell^x = \begin{pmatrix} 0 & 1 \\ 1 & 0 \end{pmatrix}, \quad \sigma_\ell^y = \begin{pmatrix} 0 & -i \\ i & 0 \end{pmatrix}, \quad \sigma_\ell^z = \begin{pmatrix} 1 & 0 \\ 0 & -1 \end{pmatrix}. \tag{2}$$

For the remainder of this discussion, we focus on the fully ferromagnetic case with $J, K > 0$. The sign of $h$ plays no role since it can be compensated by an appropriate rotation.

We define the vector operator

$$\mathbf{S} = \frac{1}{2}\sum_\ell \boldsymbol{\sigma}_\ell, \tag{3}$$

where the boldface notation $\mathbf{S} = (S^x, S^y, S^z)$ and similarly $\boldsymbol{\sigma} = (\sigma^x, \sigma^y, \sigma^z)$ denotes vector representation. Using this operator, the Hamiltonian can be expressed as

$$\mathcal{H} = -\frac{4J}{N}(S^z)^2 - 2hS^x - \frac{16K}{N^3}(S^z)^4. \tag{4}$$

In the $K \to 0$ limit, the Hamiltonian in Eq. (1) simplifies to the well-known Lipkin-Meshkov-Glick (LMG) model [47–49]. At zero temperature ($T = 0$), this system features a quantum critical point at $h = h_c = 2J$, signaling a phase transition from a paramagnetic phase, where spins align along the $x$ axis, to a ferromagnetic phase with a nonzero magnetization component along the $z$ axis. Note that while the Hamiltonian in Eq. (1) is extensive due to the rescaling of the interactions, it still remains nonadditive and therefore can accommodate ensemble inequivalence.

Hamiltonians akin to Eq. (1) have been employed to investigate a variety of physical systems under both canonical and microcanonical ensembles. In the canonical framework, the quantum critical behavior of the Dicke model [50], which describes the interaction between the motional degrees of freedom of a Bose gas and the standing wave field of an optical cavity [51,52], coincides with the one of the LMG model [53,54]. Furthermore, spin models like Eq. (1) can be experimentally realized by coupling atomic internal states to the cavity field [55–58].

Alternatively, microcanonical ensemble representations of Eq. (1) describe systems such as coupled Bose-Einstein condensates (BECs) or Bose-Hubbard models in double-well configurations [59], spin-1 BECs [60–64], and Rydberg atoms in the blockade regime [65–69]. Thus, examining the model in Eq. (1) under both canonical and microcanonical ensembles is of significant interest.

A distinctive aspect of the Hamiltonian in Eq. (1) is the incorporation of multispin interactions. While past research has primarily addressed the $K \to 0$ limit, investigating the general case aligns with current experimental attempts aimed at achieving quantum control of multibody interactions [70,71]. These interactions have already been used to model order-disorder transitions in ferroelectrics [72].

To explore the equilibrium properties of this system, we analyze it below within both the canonical and microcanonical ensembles, focusing on their respective thermodynamic potentials.

## III. MODEL ANALYSIS

In what follows, we compute the free energy and entropy of the model, starting by computing the canonical partition function $Z$ at temperature $k_B T = \beta^{-1}$ and the phase-space volume $\Omega$ with fixed energy $E$:

$$Z(\beta, J, h, K) = \mathrm{Tr}[e^{-\beta\mathcal{H}}], \tag{5}$$

$$\Omega(E, J, h, K) = \mathrm{Tr}[\delta_\Delta(E - \mathcal{H})], \tag{6}$$

where the operator $\delta_\Delta$ counts the number of eigenstates whose energies lie within the narrow window $E \in [E, E + \Delta]$. Therefore, free energy $F$ and entropy $S$ are defined as

$$F(\beta, J, h, K) = -\ln Z(\beta, J, h, K)/\beta, \tag{7}$$

$$\mathcal{S}(E, J, h, K) = \ln \Omega(E, J, h, K), \tag{8}$$

where the Boltzmann constant has been set to $k_B = 1$. Following Landau [73], Eq. (8) provides a mathematical definition of quantum microcanonical entropy. For macroscopically large systems with a nearly continuous energy spectrum, the energy window $\Delta$ must satisfy two conditions: It should be much larger than the average level spacing (which vanishes in the thermodynamic limit), yet much smaller than the total energy, which grows with the system size $N$. This ensures that the microcanonical entropy captures the fine structure of the many-body spectrum while remaining a smooth function. Although this definition initially introduces a dependence of the entropy $\mathcal{S}$ on both energy $E$ and the width $\Delta$, this dependence fades away in the thermodynamic limit. Consequently, the resulting entropy density $s = \lim_{N\to\infty} \mathcal{S}/N$ depends only on the energy density $\varepsilon = \lim_{N\to\infty} E/N$ [74].

We perform the calculation of the microcanonical entropy by considering Eq. (6) directly in the $\Delta \to 0$ limit. In this limit, we assume that the volume of the microcanonical phase space converges to the result obtained by replacing the operator $\delta_\Delta$ with the Dirac $\delta$ distribution. The validity of this procedure for Hamiltonian (1) can be confirmed *a posteriori* by observing that, when the trace in Eq. (6) is converted into an integral, the dominant contribution to the partition function arises from the saddle point, located in the region of semiclassical states with $\mathcal{S} \propto N$. Consequently, interchanging the order of the limit $\Delta \to 0$ and the trace in Eq. (6) poses no difficulty for mean-field models such as Hamiltonian (1). It is worth noting, however, that a direct computation of the partition function via the Dirac $\delta$ distribution is generally not applicable. A notable exception occurs in Hamiltonians displaying many-body localization [75].

### A. Canonical ensemble

For the canonical partition function, we get

$$\begin{aligned} Z &= \sum_{\{\vec{\tau}\}} \langle \vec{\tau}| \big(e^{-\beta H_z + \beta h \sum_i \sigma_i^x}\big) |\vec{\tau}\rangle \\ &= \lim_{N_s\to\infty} \mathrm{Tr}\Big[e^{-\frac{\beta}{N_s} H_z + \frac{\beta}{N_s} h \sum_i \sigma_i^x}\Big]^{N_s}, \end{aligned} \tag{9}$$

where $H_z$ is the part of the Hamiltonian (1) diagonal in the $z$ basis, $H_z = -\frac{4J}{N}(S^z)^2 - \frac{16K}{N^3}(S^z)^4$, and $|\vec{\tau}\rangle$ represents a generic $z$-component Ising spin configuration, $\vec{\tau} \equiv |\uparrow, \uparrow, \downarrow, \dots\rangle$, which spans a complete orthogonal basis of the total Hilbert space. In the second line of this equation, we have represented $Z$ by a product of $N_s$ Trotter slices. Introducing a closure relation in between each Trotter slice, $\mathbb{1} = \sum_{\{\vec{\tau}\}} |\vec{\tau}\rangle\langle\vec{\tau}|$, and noting that in the limit $N_s \to \infty$ one can split the exponential, we get

$$\begin{aligned} Z &= \sum_{\{\vec{\tau}(\alpha)\}} \prod_{\alpha=1}^{N_s} \langle \vec{\tau}(\alpha)| e^{-\frac{\beta}{N_s} H_z(\alpha)} e^{\frac{\beta}{N_s}\sum_i h\sigma^x(\alpha)} |\vec{\tau}(\alpha+1)\rangle \\ &= \sum_{\{\vec{\tau}(\alpha)\}} \prod_{\alpha=1}^{N_s} e^{-\frac{\beta}{N_s} E_z(\alpha)} \prod_{\alpha=1}^{N_s} \langle \vec{\tau}(\alpha)| e^{\frac{\beta}{N_s}\sum_i h\sigma^x(\alpha)} |\vec{\tau}(\alpha+1)\rangle. \end{aligned} \tag{10}$$

The index $\alpha$ in $\langle\vec{\tau}(\alpha)|$ labels the Trotter step and the energy $E_z(\alpha)$ satisfies the eigenvalue equation $H_z|\vec{\tau}(\alpha)\rangle = E_z(\alpha)|\vec{\tau}(\alpha)\rangle$. The trace in Eqs. (5) and (6) imposes periodic boundary conditions on the additional "time" direction labeled by $\alpha$, $\vec{\tau}(N_s + 1) = \vec{\tau}(1)$.

We proceed by defining the magnetization order parameter for each of the $\alpha$ slices,

$$m_z(\alpha) = \frac{1}{N}\sum_i \sigma_i^z(\alpha), \tag{11}$$

and introducing $N_s$ copies of the delta function $N\int dm_z \delta(Nm_z - \sum_i \sigma_i^z) f(Nm_z) = f(\sum_i \sigma_i^z)$. As we are working directly in the thermodynamic limit, where $N \to \infty$ and, accordingly, $\Delta \to 0$, we replace the operator $\delta_\Delta$ with the Dirac's $\delta$ distribution. We then make use of the Fourier representation of the delta function

$$\delta(x) = \frac{1}{2\pi i}\int_{-i\infty}^{+i\infty} e^{\lambda x} d\lambda. \tag{12}$$

The partition sum is then given by

$$\begin{aligned} Z &= \lim_{N_s\to\infty} \int \prod_{\alpha=1}^{N_s} \frac{dm_z(\alpha) d\lambda(\alpha)}{2\pi i N_s/(\beta N)} e^{-\frac{\beta N}{N_s}\sum_{\alpha=1}^{N_s}(e(m_z(\alpha)) + \lambda(\alpha) m_z(\alpha))} \times \sum_{\vec{\tau}(1),\dots,\vec{\tau}(N_s)} \prod_{\alpha=1}^{N_s} \langle\vec{\tau}(\alpha)| e^{\frac{\beta}{N_s}\sum_{i=1}^N [\lambda(\alpha)\sigma_i^z + h\sigma_i^x]} |\vec{\tau}(\alpha+1)\rangle \\ &= \lim_{N_s\to\infty} \int \prod_{\alpha=1}^{N_s} \frac{dm_z(\alpha) d\lambda(\alpha)}{2\pi i N_s/(\beta N)} \exp\left[-N\left(\frac{\beta}{N_s}\sum_{\alpha=1}^{N_s}(e(m_z(\alpha)) + \lambda(\alpha) m_z(\alpha)) - \ln \mathrm{Tr} \prod_{\alpha=1}^{N_s} e^{\frac{\beta}{N_s}(\lambda(\alpha)\sigma_i^z + h\sigma_i^x)}\right)\right], \end{aligned} \tag{13}$$

where we have defined $e(m_z) \equiv E_z/N = -Jm_z^2 - Km_z^4$. Note that, due to the fully connected nature of the model, the Hamiltonian is insensitive to fluctuations along the chain and only the total magnetization profile $m_z(\alpha) = \sum_{i=1}^N \sigma_i^z(\alpha)/N$ enters into the computation of the partition function. Let us note that, up to this point, no approximation has been made. We have just rewritten the partition function in terms of a path integral following the procedure of Ref. [34]. At this point, we make use of the "static" approximation that assumes that the dominant contribution to the integral comes from values of $m_z(\alpha)$ and $\lambda(\alpha)$ that are independent of $\alpha$, to obtain

$$Z \propto \int dm_z d\lambda\, \exp\left[N\beta\left(Jm_z^2 + Km_z^4 - \lambda m_z + \frac{1}{\beta}\ln\left(2\cosh\left(\beta\sqrt{\lambda^2+h^2}\right)\right)\right)\right]. \tag{14}$$

Actually, the "static approximation" is not really an approximation for the ground-state properties of Hamiltonian (4). In fact, note that Eq. (13) contains no "kinetic" term for the magnetization $m_z(\alpha)$ and the auxiliary field $\lambda(\alpha)$. Thus, there is no apparent *a priori* reason why fluctuating profiles of $m_z(\alpha)$ and $\lambda(\alpha)$ should be included in the computation of the partition sum. Moreover, it can be shown that the explicit saddle point solution of Eq. (13) yields homogeneous order parameter profiles $m_z(\alpha) = m_z$, $\lambda(\alpha) = \lambda$, independent of $\alpha$ (see Appendix C). Returning to our discussion of the microcanonical ensemble in Sec. II, we can now see how the absence of mixing between different imaginary times leads directly to a well-defined classical order parameter $m_z$. In the thermodynamic limit, it is intuitive that the transverse field and the longitudinal component of the Hamiltonian effectively commute, thereby enabling the use of $m_z$ as the order parameter.

Equation (14), and the corresponding equation (22) below for the microcanonical ensemble, can be computed by a saddle point approximation that becomes exact in the thermodynamic limit. The free energy $f = \lim_{N\to\infty} F/N$ reduces to

$$f(\beta, J, h, K) = -Jm_z^2 - Km_z^4 + \lambda m_z - \frac{1}{\beta}\ln\left(2\cosh\left(\beta\sqrt{h^2+\lambda^2}\right)\right). \tag{15}$$

In order to apply the saddle point method, we deform the contour of integration of $\lambda$ from $(-i\infty, +i\infty)$ to $(-i\infty + a, +i\infty + a)$, where $a$ is a real number, chosen such that the new contour passes through the saddle point of the free energy $f$. The saddle point condition imposes the free energy to be an extremum with respect to the variational parameters $m_z$ and $\lambda$:

$$\frac{\partial f}{\partial \lambda} = 0: \quad m_z = \frac{\lambda}{\sqrt{\lambda^2+h^2}}\tanh(\beta\sqrt{\lambda^2+h^2}), \tag{16}$$

$$\frac{\partial f}{\partial m_z} = 0: \quad \lambda = 2Jm_z + 4Km_z^3. \tag{17}$$

For later convenience, we write the free energy in terms of a single order parameter $m_z$:

$$f(\beta, J, h, K) = Jm_z^2 + 3Km_z^4 - \frac{1}{\beta}\ln\left(2\cosh\left(\beta\sqrt{h^2 + (2Jm_z + 4Km_z^3)^2}\right)\right). \tag{18}$$

In Sec. IV, we analyze this free energy to obtain the canonical phase diagram.

### B. Microcanonical ensemble

Following the reasoning presented in Ref. [76], the microcanonical phase-spece volume, for a sufficiently large system, can be obtained by treating the energy spectrum as continuous and applying the representation of the Dirac $\delta$ distribution given in Eq. (12):

$$\Omega = \int_{-i\infty+a}^{+i\infty+a} \frac{d\gamma}{2\pi i} \sum_{\{\vec{\tau}\}} \langle \vec{\tau} | (e^{\gamma(E - H_z + h\sum_i \sigma_i^x)}) | \vec{\tau} \rangle. \tag{19}$$

By following the same procedure described for the canonical ensemble, we can rewrite the phase-space volume $\Omega$ as

$$\Omega = \int \frac{d\gamma}{2\pi i} \sum_{\{\vec{\sigma}(\alpha)\}} \prod_{\alpha=1}^{N_s} e^{\gamma(E - E_z(\alpha))/N_s} \times \prod_{\alpha=1}^{N_s} \langle \vec{\tau}(\alpha) | e^{\frac{\gamma h}{N_s}\sum_i \sigma^x(\alpha)} | \vec{\tau}(\alpha+1) \rangle. \tag{20}$$

It is important to note that, while in the canonical expression $\beta$ is a fixed parameter, here the corresponding parameter, $\gamma$, is being integrated over the entire range $[-\infty, \infty]$. Therefore, even though it was handy to introduce the Dirac representations scaled with a $\beta$ factor in the canonical derivation, we cannot naively proceed in the same manner for the computation of $\Omega$. Taking this fact into account, we instead introduce the unscaled Dirac representation as

$$\Omega = \lim_{N_s\to\infty} \int \frac{d\gamma}{2\pi i} \int \prod_{\alpha=1}^{N_s} \frac{dm_z(\alpha) d\lambda(\alpha)}{2\pi N_s} \exp\left[N\left(\frac{1}{N_s} \times \sum_{\alpha=1}^{N_s} (\gamma(\varepsilon - e(m_z(\alpha))) - \lambda(\alpha) m_z(\alpha)) + \ln \mathrm{Tr} \prod_{\alpha=1}^{N_s} e^{\frac{1}{N_s}(\lambda(\alpha)\sigma_i^z + \gamma h \sigma_i^x)}\right)\right], \tag{21}$$

where $\varepsilon \equiv E/N$ is the fixed total energy per spin. In the same spirit as above, the dominant contribution comes from values of $m_z(\alpha)$ and $\lambda(\alpha)$ that are independent of $\alpha$:

$$\Omega \propto \int d\gamma dm_z d\lambda\, \exp\left[N\left(\gamma\left(Jm_z^2 + Km_z^4 + \varepsilon\right) - \lambda m_z + \ln 2\cosh(\sqrt{\lambda^2 + \gamma^2 h^2})\right)\right]. \tag{22}$$

We observe that, in the microcanonical case, we have an additional quantity $\gamma$, which accounts for the fact that energy is fixed. Again, due to the factor $N$ in the exponential, the entropy $\mathcal{S} \equiv \ln \Omega / N$ is completely dominated by values of $\gamma, \lambda, m_z$ that maximize the entropy

$$\mathcal{S}(\varepsilon, J, h, K) = \gamma\left(Jm_z^2 + Km_z^4 + \varepsilon\right) - \lambda m_z + \ln 2\cosh(\sqrt{\lambda^2 + \gamma^2 h^2}), \tag{23}$$

with

$$\partial_\lambda \mathcal{S} = 0: m_z = \frac{\lambda \tanh(\sqrt{\lambda^2 + \gamma^2 h^2})}{\sqrt{\lambda^2 + \gamma^2 h^2}}, \tag{24}$$

$$\partial_{m_z} \mathcal{S} = 0: \frac{\lambda}{\gamma} = 2Jm_z + 4Km_z^3, \tag{25}$$

$$\partial_\gamma \mathcal{S} = 0: \ \varepsilon = -Jm_z^2 - Km_z^4 - \frac{\gamma h^2}{\sqrt{\lambda^2+\gamma^2h^2}} \tanh(\sqrt{\lambda^2+\gamma^2h^2}). \quad (26)$$

Here, we have chosen the arbitrary parameter $a$ such that the $\gamma$ contour of integration passes through the saddle point. Substituting these relations into Eq. (23), and following the steps outlined at the beginning of Appendix A, one can rewrite the entropy as a function of the magnetization $m_z$ and energy $\varepsilon$:

$$\mathcal{S}(\varepsilon, J, h, K) = -\operatorname{arctanh}\left(\sqrt{m_z^2 + \frac{(\varepsilon + Jm_z^2 + Km_z^4)^2}{h^2}}\right) \times \sqrt{m_z^2 + \frac{(\varepsilon + Jm_z^2 + Km_z^4)^2}{h^2}} + \ln\left(\frac{2h}{\sqrt{h^2 - m_z^2h^2 - (\varepsilon + Jm_z^2 + Km_z^4)^2}}\right). \quad (27)$$

## IV. THE PHASE DIAGRAM

In the following, we analyze the finite-temperature phase diagrams in both ensembles.

### A. Canonical ensemble

Let us begin by determining the location of the second-order phase transition line and the tricritical point. We start by expanding Eq. (18) in terms of $m_z^2$:

$$f(\beta, J, h, K) \approx f_0 + b_2 m_z^2 + b_4 m_z^4 + \mathcal{O}(m_z^6), \quad (28)$$

where

$$b_2 = J - 2J^2 \frac{\tanh(\beta h)}{h}, \quad (29)$$

$$b_4 = 3K - \frac{2J}{h^3}\big(\beta h J^3 \cosh^{-2}(\beta h) + (4h^2K - J^3)\tanh(\beta h)\big). \quad (30)$$

Setting $b_2 = 0$ yields the second-order transition line

$$\frac{h_c}{2J} = \tanh(\beta h_c). \quad (31)$$

Setting both $b_2 = 0$ and $b_4 = 0$ determines the tricritical point

$$K_{\text{tcp}} = \frac{J^3}{h_c^2} + \frac{\beta J^2}{2}\left(1 - \frac{4J^2}{h_c^2}\right). \quad (32)$$

Expanding near low temperatures, we obtain

$$b_2 = 0: \quad h_c[\text{CE}] = 2J(1 - 2e^{-4\beta J}), \quad (33)$$

$$b_4 = 0: \quad K_{\text{tcp}}[\text{CE}] = \frac{J}{4} - 2\beta J^2 e^{-4\beta J}, \quad (34)$$

which coincides with the result reported in Ref. [33]. It is worthwhile noting that the critical field $h/J$ does not depend on $K/J$, but is only a function of the temperature.

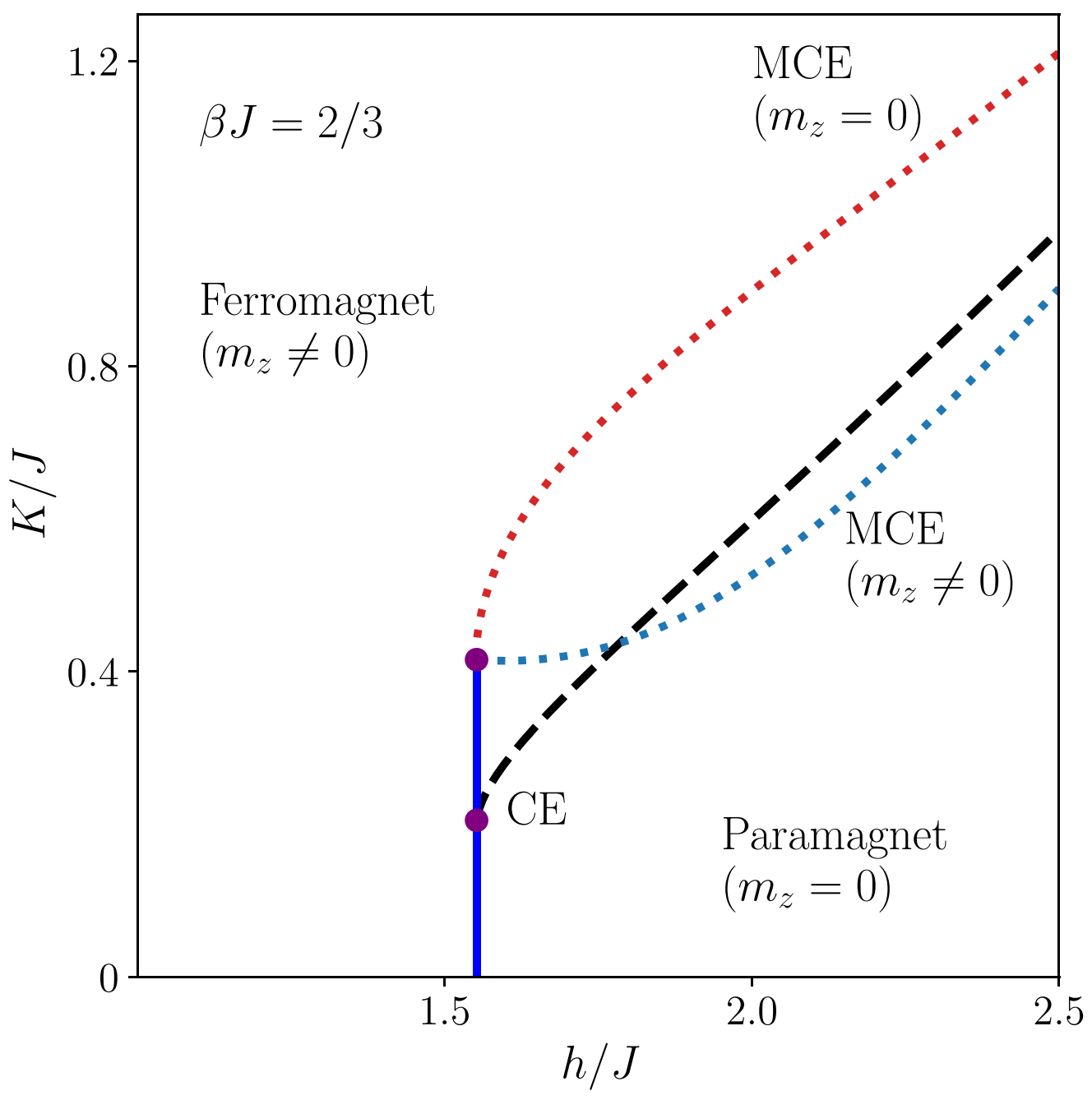

FIG. 1. The canonical and microcanonical $(h/J, K/J)$ phase diagrams at a given temperature ($\beta J = 2/3$) are shown. The microcanonical critical line (solid blue) coincides with the canonical one but extends beyond the canonical tricritical point (CE). The two microcanonical first-order transition lines (dotted) correspond, respectively, to either the $m_z = 0$ solution (dotted red) or to the spontaneously magnetized state $m_z = m_z^*$ (dotted blue). The canonical first-order transition line is depicted by the black dashed line.

The first-order transition line is determined by finding numerically for which values of $(K/J, h/J)$ the system undergoes a discontinuous change to a state with nonvanishing magnetization $m_z^*$. This happens when $f(m_z = 0)$ and $f(m_z = m_z^*)$ are both global minima of the free energy, i.e.,

$$f(m_z = 0) = f(m_z^*), \quad \partial_{m_z} f(m_z)|_{m_z=0} = \partial_{m_z} f(m_z)|_{m_z=m_z^*} = 0, \quad (35)$$

yielding the following conditions:

$$m_z^* = \frac{\tanh\left(\beta\sqrt{h^2 + (2Jm_z^* + 4Km_z^{*3})^2}\right)}{\sqrt{h^2 + (2Jm_z^* + 4Km_z^{*3})^2}}(2Jm_z^* + 4Km_z^{*3}),$$

$$Jm_z^{*2} + 3Km_z^{*4} + \frac{1}{\beta}\ln(\cosh\beta h) = \frac{1}{\beta}\ln\left(\cosh\left(\beta\sqrt{h^2 + (2Jm^* + 4Km^{*3})^2}\right)\right), \quad (36)$$

whose solution is found numerically and reported in Fig. 1.

### B. Microcanonical ensemble

To determine the second-order phase transition line and the tricritical point in the microcanonical ensemble, we expand Eq. (27) in terms of $m_z^2$ and again locate for which conditions the coefficients of the expansion vanish.

$$\mathcal{S}(\varepsilon, J, h, K) \approx s_0 + a_2 m_z^2 + a_4 m_z^4 + \mathcal{O}(m^6), \quad (37)$$

with

$$a_2 = -\frac{(h^2 + 2\varepsilon J)\text{arctanh}(\varepsilon/h)}{2h\varepsilon}, \tag{38}$$

$$a_4 = \frac{h^2 + 2\varepsilon J}{8\varepsilon^2(\varepsilon^2 - h^2)} - \frac{h^4 + 4\varepsilon h^2 J - 8\varepsilon^3 K}{8h\varepsilon^3}\text{arctanh}(\varepsilon/h). \tag{39}$$

The resulting critical line is found by imposing

$$h^2 + 2\varepsilon J = 0, \tag{40}$$

which, together with

$$h^4 + 4\varepsilon h^2 J - 8\varepsilon^3 K = 0, \tag{41}$$

yields the tricritical point. In order to proceed and compare with the canonical ensemble, one has to express $\varepsilon$ in terms of the temperature. Along the critical line, $m_z = 0$, the temperature is in direct connection with the energy,

$$\beta = \frac{\partial \mathcal{S}}{\partial \varepsilon} = -\frac{\text{arctanh}(\varepsilon/h)}{h}, \tag{42}$$

which gives

$$\varepsilon = -h \tanh \beta h. \tag{43}$$

Inserting expression (43) in (40), the microcanonical ([MCE]) critical line can be found by numerically solving

$$h_c[\text{MCE}] = 2J \tanh(\beta h_c), \tag{44}$$

which coincides with the result in the canonical ensemble (31). Along the critical line, Eq. (41) yields the tricritical point at

$$K_{\text{tcp}}[\text{MCE}] = \frac{J}{4 \tanh(\beta h_c)^2}. \tag{45}$$

This expression, together with the respective result in the canonical ensemble (32), demonstrate the inequivalence of the two ensembles. While both ensembles yield the same expression for the critical lines [Eqs. (31) and (44)], their tricritical points differ. At a fixed temperature, the canonical tricritical point, given by Eq. (34), occurs at a lower value of $K/J$ than the corresponding microcanonical tricritical point, as shown in Fig. 1, which presents the $(h/J, K/J)$ phase diagram at $\beta J = 2/3$. To determine the first-order transition lines, we have to maximize the entropy with respect to the order parameter $m_z$ at fixed energy. To compare with the canonical solution, one needs to fix the temperature instead of the energy. In order to do so, we have to relate the energy to the temperature. However, for systems that present regions with negative specific heat, there is no one-to-one correspondence between energy and temperature. In this case, we can determine the temperature either by looking at the solution in the paramagnetic regime ($m_z = 0$) or at the solution in the ferromagnetic regime ($m_z = m_z^* \neq 0$), given, respectively, by

$$\beta = -\frac{\text{arctanh}(\varepsilon/h)}{h}, \tag{46}$$

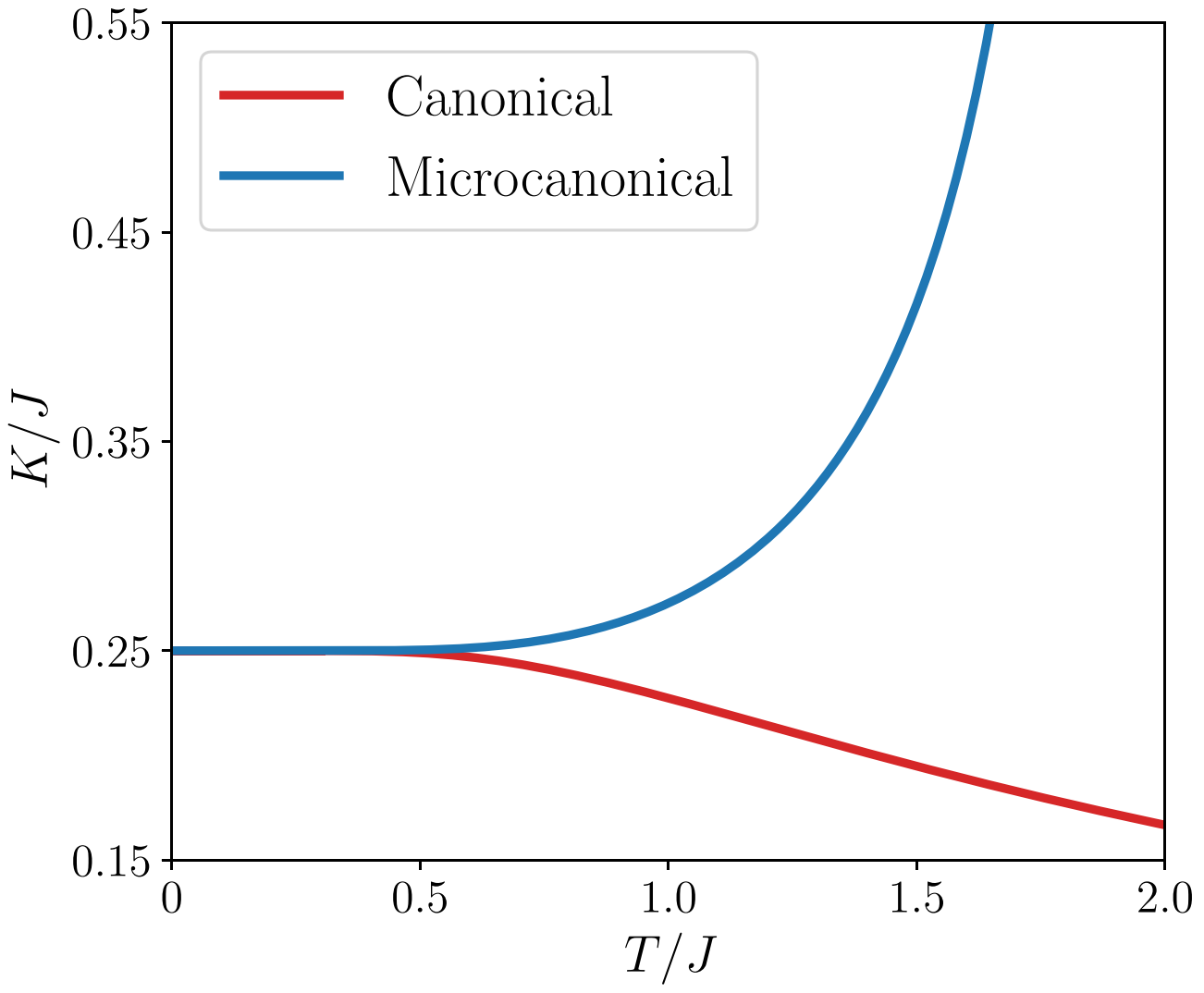

FIG. 2. Tricritical point $K/J$ against $T/J$ in the canonical and microcanonical ensembles. Note that the field $h/J$ varies along the lines [see Eqs. (31) and (44)].

$$\beta = -\frac{\left(\varepsilon + Jm_z^{*2} + Km_z^{*4}\right)}{h\sqrt{h^2 m_z^{*2} + \left(\varepsilon + Jm_z^{*2} + Km_z^{*4}\right)^2}} \times \text{arctanh}\left(\sqrt{m_z^{*2} + \frac{\left(\varepsilon + Jm_z^{*2} + Km_z^{*4}\right)^2}{h^2}}\right). \tag{47}$$

Similarly to the canonical ensemble, in order to determine the first-order transition line, we require the entropy $\mathcal{S}$ to have three global maxima at $m_z = 0, \pm m_z^*$, i.e.,

$$h^2 + \left(\varepsilon + Jm_z^{*2} + Km_z^{*4}\right)\left(2J + 4Km_z^{*2}\right) = 0,$$
$$h^2 m_z^{*2} + \left(\varepsilon + Jm_z^{*2} + Km_z^{*4}\right)^2 = \varepsilon^2. \tag{48}$$

The first equation expresses the requirement that the solution $m_z = \pm m_z^*$ is a local extremum of the entropy, while the second equation results from the requirement that the entropies at $m_z = 0$ and $\pm m_z^*$ are equal. For given $(J, h, K)$, these two equations are solved for $\epsilon$ and $m_z^*$, yielding the energy and the magnetization in the ordered state. Using these values in Eqs. (46) and (47) to determine the two temperatures of the coexisting states gives rise to the two first-order transition lines reported in Fig. 1. In combination with the canonical transition line, these two microcanonical lines complete the phase diagram that was previously only schematically reported in Ref. [33].

In Fig. 2, we display the tricritical coupling $K/J$ for both ensembles [Eqs. (34) and (45)] as a function of $T/J$. At $T = 0$, the tricritical points of the two ensembles coincide, but as the temperature increases, the microcanonical tricritical point varies toward larger values of $K/J$, while the canonical tricritical point varies more gradually toward smaller values of $K/J$. It is important to note that, while the magnetic field $h/J$ varies along the two lines, its value at any given temperature is the same on both lines.

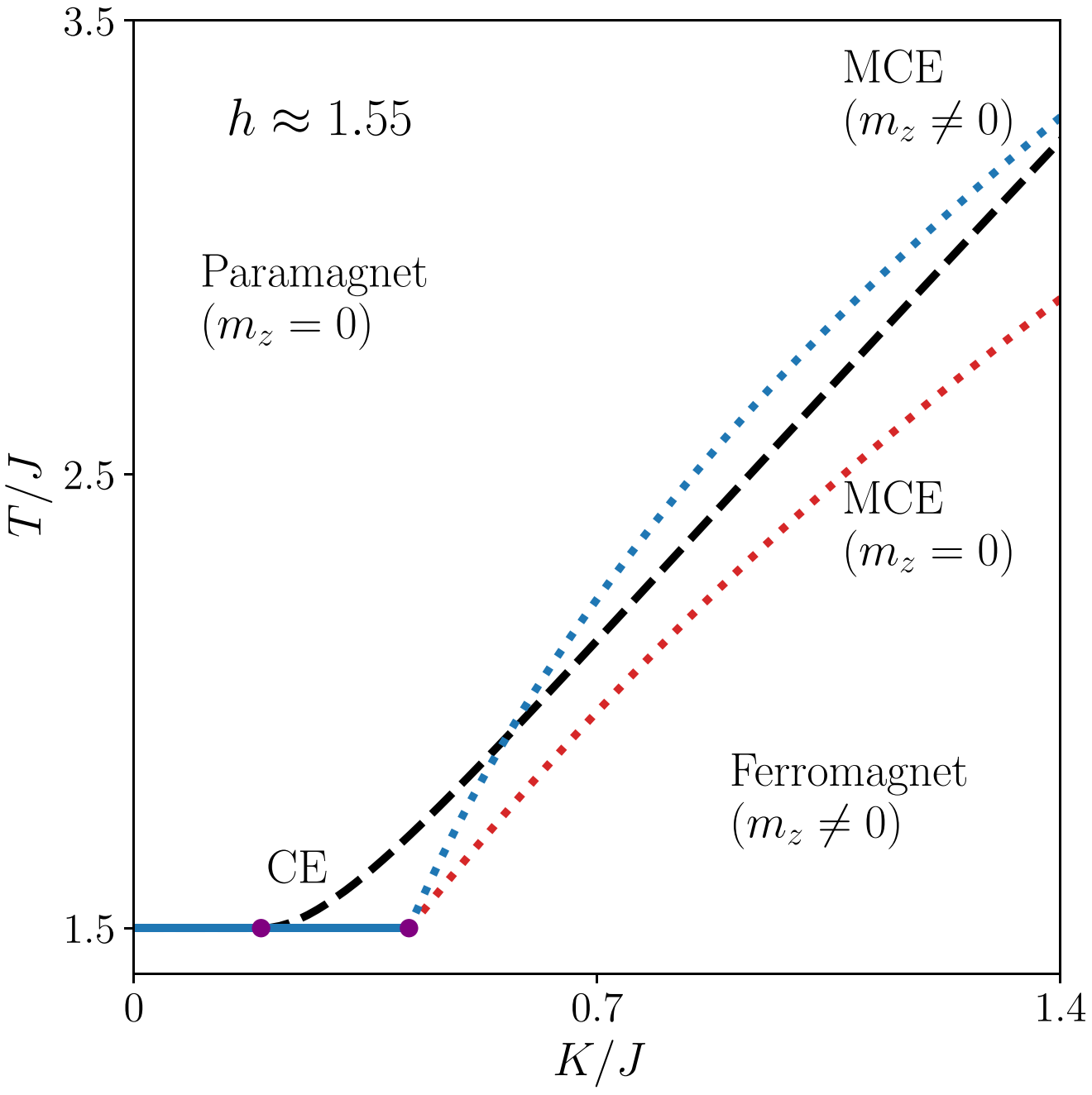

FIG. 3. The canonical and microcanonical $(K/J, T/J)$ phase diagrams at a given $h/J \approx 1.55$, chosen such that the critical line and the two tricritical points are located at $\beta J = 2/3$. As in Fig. 1, the two microcanonical first-order lines (dotted) correspond, respectively, to either the $m_z = 0$ solution or to the spontaneously magnetized $m_z = m_z^*$ one.

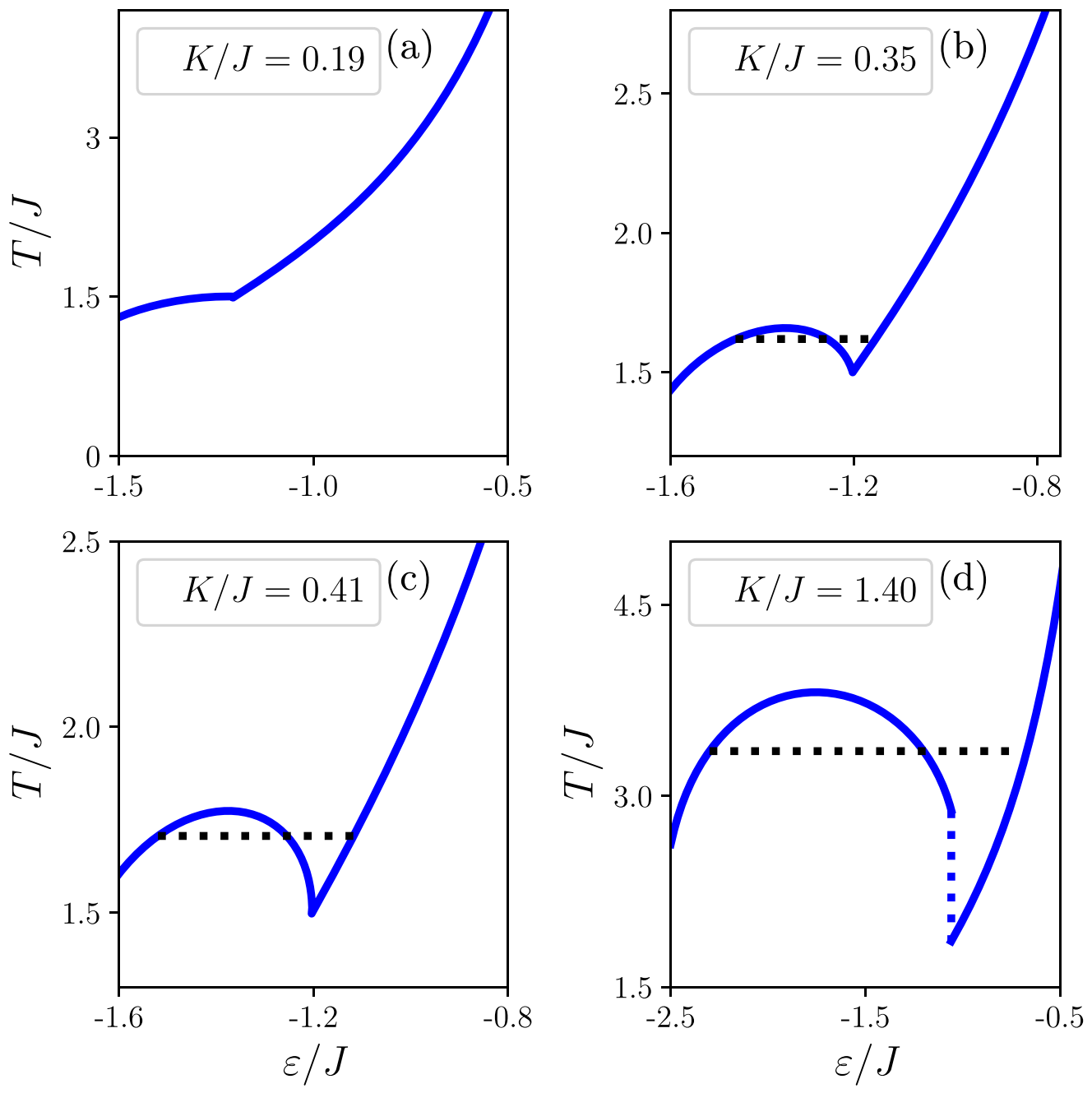

FIG. 4. Temperature versus energy relation in the microcanonical ensemble for $h/J \approx 1.55$ and increasing values of $K/J$. Near the canonical tri-critical point (panel (a) with $T = 1/5$), the lower energy branch of the caloric curve has zero slope at the intersection point with the higher energy branch. The horizontal dotted line in panels (b)–(d) is the Maxwell construction in the canonical ensemble, which identifies the canonical first-order transition temperature. The dotted vertical line in panel (d) shows the *"temperature jump"* in the microcanonical ensemble.

## V. CALORIC CURVES

To complete and complement the study of the phase diagram in the microcanonical ensemble, we display in Fig. 3 the $(T/J, K/J)$ phase diagram for a given field $h/J$ and in Fig. 4 the temperature-energy relation $T(\varepsilon)$.

As in Fig. 1, we see in Fig. 3 the two distinct tricritical points of the canonical (CE) and microcanonical (MCE) ensembles. The first-order transition line in the canonical ensemble (dashed) differs from the two lines in the microcanonical ensemble (dotted).

In Fig. 4, we display the caloric curves for $h/J \approx 1.55$ and several values of $K/J$. We observe that the critical temperature in both ensembles is independent of $K/J$ and depends only on $h/J$. The value of $h/J$ is arbitrarily chosen such that the temperature on the critical line and tricritical point in both ensembles is $\beta J = 2/3$. Each caloric curve is composed of two branches, a low-energy branch corresponding to the spontaneously magnetized state $m_z^*$ and a high-energy one for the paramagnetic phase $m_z = 0$.

The two branches intersect where their respective entropies become equal. For $K/J \approx 0.19$, corresponding to the canonical tricritical point at $\beta J = 2/3$, see Eqs. (31) and (32). The lower energy branch of the caloric curve has zero slope at the intersection point with the higher energy branch [see Fig. 4(a)]. At this point, the specific heat of the magnetized phase diverges. For $K/J = 0.35$ [see Fig. 4(b)], where the system lies in between the canonical and the microcanonical tricritical points, an energy domain with *negative specific heat* in the microcanonical ensemble, $\partial T/\partial \epsilon < 0$, is present. Increasing $K/J$ to $K/J \approx 0.41$ [see Fig. 4(c)], the microcanonical tricritical point is reached [see Eqs. (44) and (45)]. Here, the slope of the lower energy branch of the caloric curve diverges. At higher values of $K/J$, a discontinuity develops [see Fig. 4(d)] for $K/J = 1.40$), signaling a *"temperature jump"* at the transition.

## VI. CONCLUSIONS

In this paper, we extend a previous study of the phase diagram of a model with long-range and multispin interactions. We provide an alternative analytical derivation of its canonical free energy and microcanonical entropy using path integral methods, and accurately characterize its various phases of matter and different transition lines.

At $T = 0$, the phase diagram is solely determined by the ground-state properties, resulting in identical phase diagrams for both ensembles. However, at finite temperatures, the phase diagrams are different in the two ensembles. Specifically, we found that the tricritical point shifts upward in the microcanonical description and downward in the canonical description as the temperature increases. We also investigate the relationship between temperature $T$ and energy $\varepsilon$ in the microcanonical case, demonstrating that, for certain parameters, the model exhibits *negative specific heat* and *"temperature jumps"* near the first-order phase transition.

To understand the robustness of this effect, it would be valuable to explore the validity of these results in other quantum models, such as spin systems with long-range interactions where the interaction strength decays as a power law with the distance.

As the field of AMO continues to advance, we anticipate that these results could be experimentally tested on quantum platforms, particularly in the context of cold atoms with cavity-mediated interactions. Specifically, the Hamiltonian in Eq. (1) for $K = 0$ is the paradigmatic LMG model, which has already been used to describe certain cavity QED platforms [43,77,78]. Realizing the four-body interaction term at $K > 0$ may be achieved by leveraging recent findings on cavity-mediated pair creation [79].

## ACKNOWLEDGMENTS

This research was supported in part by NSF Grant No. PHY-1748958 to the Kavli Institute for Theoretical Physics (KITP). This research was funded by the Swiss National Science Foundation (SNSF) Grant No. 200021 207537, by the Deutsche Forschungsgemeinschaft (DFG, German Research Foundation) under Germany's Excellence Strategy EXC2181/1-390900948 (the Heidelberg STRUCTURES Excellence Cluster), the Swiss State Secretariat for Education, Research and Innovation (SERI), and by the Center for Scientific Excellence at the Weizmann Institute of Science. S.R. acknowledges support from the MUR PRIN2022 project BECQuMB Grant No. 20222BHC9Z.

## DATA AVAILABILITY

No data were created or analyzed in this study.

## APPENDIX A: EQUIVALENCE BETWEEN METHODS

In a previous study [33], a different approach was applied to calculate the free energy and the entropy of the model. To compute the thermodynamic potentials, the authors of Ref. [33] decompose the Hilbert space into different total spin sectors. They use the formulas below for the partition sum and the microcanonical volume, respectively,

$$Z(\beta, J, h, K) = \sum_S g(S) \sum_{S^z=-S}^{S} \langle S, S^z | e^{-\beta\mathcal{H}} | S, S^z\rangle, \tag{A1}$$

$$\Omega(E, J, h, K) = \sum_S g(S) \sum_{S^z=-S}^{S} \langle S, S^z | \delta(E - \mathcal{H}) | S, S^z\rangle. \tag{A2}$$

Here, $S$ labels the total quantum spin (composed of $N$ tensor products of $1/2$ spins) and $S_z$ the magnetization along the $z$ direction. The factor $g(S)$ accounts for the degeneracy that comes from the multiple ways to arrange the microscopic $1/2$ spins in order to form a total spin $S$ [72]. For the large spin $S$ sectors, one can approximate the quantum partition function by a classical integral over the surface of a sphere with radius $S = M$ and parametrized such that $\mathbf{S} = Ms\mathbf{m}$, where $\mathbf{m} \equiv (m_x, m_y, m_z) = (\sin\theta\cos\phi, \sin\theta\sin\phi, \cos\theta)$ [80]. Moreover, in the thermodynamic limit, one can approximate the sum over $S$ in Eqs. (A1) and (A2) by an integral over the continuous variable of total spin $s$. Altogether and because of the mean-field nature of the problem, everything is dominated by a saddle point that enforces $\phi = 0$ and depends only on two variational parameters $s$ and $m_z$. This gives, in the thermodynamic limit, the following free energy:

$$f(\beta, J, h, K) = \varepsilon - \mathcal{S}/\beta = -Js^2 m_z^2 - Ks^4 m_z^4 - hs\sqrt{1 - m_z^2} + \frac{1}{\beta}\left[\frac{1+s}{2}\ln\frac{1+s}{2} + \frac{1-s}{2}\ln\frac{1-s}{2}\right], \tag{A3}$$

and entropy

$$\mathcal{S} = -\frac{1+s}{2}\ln\left(\frac{1+s}{2}\right) - \frac{1-s}{2}\ln\left(\frac{1-s}{2}\right). \tag{A4}$$

Note that the entropy in Eq. (A4) exactly corresponds to the logarithm of $g(S)$ for large $N$.

In what follows, we show how, even though *a priori* different, Eqs. (A3) and (A4) give rise to the same phase diagrams as Eqs. (18) and (27). We start by guiding the reader through the steps taken to derive Eq. (27) from Eq. (23). In order to find an expression for $\gamma$, let us recall Eq. (26) for the energy

$$\varepsilon = -Jm_z^2 - Km_z^4 - \frac{h^2}{\sqrt{(\lambda/\gamma)^2 + h^2}}\tanh\left(\gamma\sqrt{(\lambda/\gamma)^2 + h^2}\right), \tag{A5}$$

and the relation in Eq. (25),

$$\frac{\lambda}{\gamma} = 2Jm_z + 4Km_z^3. \tag{A6}$$

We can then substitute the $(Jm_z^2 + Km_z^4 + \varepsilon)$ and $m_z$ terms in the entropy reported in Eq. (23) to obtain

$$\mathcal{S}(\varepsilon, J, h, K) = -\sqrt{\lambda^2 + \gamma^2 h^2}\tanh\sqrt{\lambda^2 + \gamma^2 h^2} + \ln\left(2\cosh\left(\sqrt{\lambda^2 + \gamma^2 h^2}\right)\right). \tag{A7}$$

Moreover, by inverting the hyperbolic tangent in Eq. (A5) and substituting $\lambda/\gamma$ with Eq. (A6), we find

$$\gamma = -\operatorname{arctanh}\left(\frac{\varepsilon + Jm_z^2 + Km_z^4}{h^2}\sqrt{h^2 + \left(2Jm_z + 4Km_z^3\right)^2}\right) \times \frac{1}{\sqrt{h^2 + \left(2Jm_z + 4Km_z^3\right)^2}}. \tag{A8}$$

Let us also note that by rewriting Eq. (26) as

$$\varepsilon + Jm_z^2 + Km_z^4 = -\frac{\gamma}{\lambda}\frac{h^2}{\sqrt{1 + h^2(\gamma/\lambda)^2}}\tanh\left(\sqrt{\lambda^2 + \gamma^2 h^2}\right), \tag{A9}$$

and making use of Eqs. (24) and (25), we can relate $m_z$ and $\varepsilon$ as follows:

$$h^2 = -\left(2J + 4Km_z^2\right)\left(\varepsilon + Jm_z^2 + Km_z^4\right). \tag{A10}$$

This equation and the fact that $J, K \geqslant 0$ imply that $\varepsilon + Jm_z^2 + Km_z^4 \leqslant 0$. It allows us to rewrite the expression for $\gamma$ in Eq. (A8) as

$$\gamma = \operatorname{arctanh}\left(\sqrt{m_z^2 + \frac{\left(\varepsilon + Jm_z^2 + Km_z^4\right)^2}{h^2}}\right) \times \frac{1}{\sqrt{h^2 + \left(2Jm_z + 4Km_z^3\right)^2}}. \tag{A11}$$

Equivalently, making use of Eq. (25), this equation may be expressed as

$$\tanh(\sqrt{\lambda^2+\gamma^2 h^2}) = \sqrt{m_z^2 + \frac{\left(\varepsilon + J m_z^2 + K m_z^4\right)^2}{h^2}}, \tag{A12}$$

which, in combination with the expression for the entropy given in Eq. (A7) and using the trigonometric relation

$$\cosh\left(\operatorname{arctanh}(x)\right) = \frac{1}{\sqrt{1-x^2}}, \tag{A13}$$

leads to the expression given in Eq. (27) for the entropy.

In order to prove the equivalence between the two approaches, let us observe that minimizing the free energy [Eq. (A3)] with respect to the parameters $m_z$ and $s$ gives the following constraints:

$$2Js + 4Ks^3 m_z^2 = \frac{h}{\sqrt{1-m_z^2}}, \tag{A14}$$

$$\frac{1}{\beta}\operatorname{arctanh}(s) = 2Jsm_z^2 + 4Ks^3 m_z^4 + h\sqrt{1-m_z^2}. \tag{A15}$$

Plugging them back into Eq. (A3) allows us to rewrite the free energy as

$$f(\beta, J, h, K) = Js^2 m_z^2 + 3Ks^4 m_z^4 - \frac{s}{\beta}\operatorname{arctanh}(s) + \frac{1}{\beta}\left[\frac{1+s}{2}\ln\frac{1+s}{2} + \frac{1-s}{2}\ln\frac{1-s}{2}\right]. \tag{A16}$$

Next, we make use of the following identity:

$$\frac{1+s}{2}\ln\frac{1+s}{2} + \frac{1-s}{2}\ln\frac{1-s}{2} = \ln\frac{\sqrt{1-s^2}}{2} + s\operatorname{arctanh}(s), \tag{A17}$$

and Eq. (A15) is finally rewritten as

$$f(\beta, J, h, K) = Js^2 m_z^2 + 3Ks^4 m_z^4 - \frac{1}{\beta}\ln\left(2\cosh\left(\beta\sqrt{h^2 + \left(2Jsm_z + 4Ks^3 m_z^3\right)^2}\right)\right), \tag{A18}$$

where we have used the fact that $1-\tanh(s)^2 = \operatorname{sech}(s)^2$ and combined Eqs. (A14) and (A15) to notice that

$$s = \tanh\left(\beta\sqrt{h^2 + \left(2Jsm_z + 4Ks^3 m_z^3\right)^2}\right).$$

This proves the equivalence between the results when rescaling the magnetization $m_z \equiv sm_z$. The equivalence can also be shown for the microcanonical ensemble by looking back at Eq. (24) and identifying

$$s = \tanh(\sqrt{\lambda^2+\gamma^2 h^2}), \quad m_z = \frac{\lambda}{\sqrt{\lambda^2+\gamma^2 h^2}}. \tag{A19}$$

Rewriting Eq. (27) in terms of $\gamma$ and $\lambda$,

$$\mathcal{S}(J, h, K) = -\sqrt{\lambda^2+\gamma^2 h^2}\,\operatorname{arctanh}(\sqrt{\lambda^2+\gamma^2 h^2}) + \ln\frac{1}{\sqrt{1-(\lambda^2+\gamma^2 h^2)}}. \tag{A20}$$

Looking back at Eqs. (A17) and (A19), it is straightforward to see the equivalence with Eqs. (A20) and (A4).

## APPENDIX B: MAXIMIZATION OF THE ENTROPY

In this Appendix, we show, in a straightforward way, how the maximization of the entropy that leads to Eq. (48) is obtained. We begin by pointing out how the entropy in Eq. (27) depends only on a single argument

$$\mathcal{S}(J,h,K) = -\sqrt{x}\,\operatorname{arctanh}(\sqrt{x}) + \ln\left(\frac{2}{\sqrt{1-x}}\right), \tag{B1}$$

where

$$x \equiv m_z^2 + \frac{\left(\varepsilon + Jm_z^2 + Km_z^4\right)^2}{h^2}. \tag{B2}$$

The entropy in Eq. (B1) is monotonically decreasing in the interval $x \in [0, 1]$ where the entropy is defined. Therefore, obtaining the maximum with respect to $m_z$ is equivalent to

$$\left.\frac{\partial x}{\partial m_z}\right|_{m=m_z^*} = 0. \tag{B3}$$

Requiring that both entropies at $m_z = 0$ and $m_z = m_z^*$ are equal is analog to solving

$$x|_{m_z=0} = x|_{m_z=m_z^*}. \tag{B4}$$

Equations (B3) and (B4) are exactly Eqs. (48) and (27).

## APPENDIX C: SUPPRESSION OF FLUCTUATIONS IN THE SADDLE POINT SOLUTION

Let us demonstrate that the only path that contributes in the thermodynamic limit to the canonical partition sum (13) is the one where $m_z(\alpha)$ and $\lambda(\alpha)$ are $\alpha$ independent. The extension to the microcanonical phase-space volume is straightforward. Since the exponential in Eq. (13) only contains on-site terms, the trace can be performed to obtain

$$Z = \lim_{N_s\to\infty}\int\prod_{\alpha=1}^{N_s}\frac{dm_z(\alpha)d\lambda(\alpha)}{2\pi i N_s/(\beta N)}\exp\left[-N\left(\frac{\beta}{N_s}\sum_{\alpha=1}^{N_s}(e(m_z(\alpha)) + \lambda(\alpha)m_z(\alpha)) - \ln\operatorname{Tr}\prod_{\alpha=1}^{N_s} e^{\frac{\beta}{N_s}(\lambda(\alpha)\sigma_i^z + h\sigma_i^x)}\right)\right], \tag{C1}$$

$$= \lim_{N_s\to\infty}\int\prod_{\alpha=1}^{N_s}\frac{dm_z(\alpha)d\lambda(\alpha)}{2\pi i N_s/(\beta N)}\exp\left[-N\left(\frac{\beta}{N_s}\sum_{\alpha=1}^{N_s}(e(m_z(\alpha)) + \lambda(\alpha)m_z(\alpha)) - \ln 2\cosh\left(\sqrt{\left(\frac{\beta}{N_s}\sum_{\alpha=1}^{N_s}\lambda(\alpha)\right)^2 + (\beta h)^2}\right)\right)\right]. \tag{C2}$$

To go from Eq. (C1) to Eq. (C2), we have implicitly undone the Trotter decomposition. In order to show explicitly how to do so, let us unwind up to Eq. (13) in the article:

$$Z = \lim_{N_s\to\infty} \int \prod_{\alpha=1}^{N_s} \frac{dm_z(\alpha)d\lambda(\alpha)}{2\pi i N_s/(\beta N)} e^{-\frac{\beta N}{N_s}\sum_{\alpha=1}^{N_s}(e(m_z(\alpha))+\lambda(\alpha)m_z(\alpha))} \sum_{\vec{\tau}(1),\ldots,\vec{\tau}(N_s)} \prod_{\alpha=1}^{N_s} \langle\vec{\tau}(\alpha)| e^{\frac{\beta}{N_s}\sum_{i=1}^{N}[\lambda(\alpha)\sigma_i^z+h\sigma_i^x]}|\vec{\tau}(\alpha+1)\rangle, \tag{C3}$$

and realize that while $\lambda(\alpha)$ depends on the Trotter step, it does not depend on the spin configuration $\vec{\tau}(\alpha)$. We can then "undo" the Trotter decomposition by identifying a copy of the identity in between each Trotter slice

$$\begin{aligned}\sum_{\vec{\tau}(1),\ldots,\vec{\tau}(N_s)} \prod_{\alpha=1}^{N_s} \langle\vec{\tau}(\alpha)| e^{\frac{\beta}{N_s}\sum_{i=1}^{N}[\lambda(\alpha)\sigma_i^z+h\sigma_i^x]}|\vec{\tau}(\alpha+1)\rangle &= \sum_{\vec{\tau}(1)}\langle\vec{\tau}(\alpha=1)| e^{\frac{\beta}{N_s}\sum_{i=1}^{N}[\lambda(1)\sigma_i^z+h\sigma_i^x]}\left\{\sum_{\vec{\tau}(\alpha=2)}|\vec{\tau}(\alpha=2)\rangle\langle\vec{\tau}(\alpha=2)|\right\}\\
&\times e^{\frac{\beta}{N_s}\sum_{i=1}^{N}[\lambda(2)\sigma_i^z+h\sigma_i^x]}\left\{\sum_{\vec{\tau}(\alpha=2)}|\vec{\tau}(\alpha=3)\rangle\langle\vec{\tau}(\alpha=3)|\right\}\\
&\times e^{\frac{\beta}{N_s}\sum_{i=1}^{N}[\lambda(3)\sigma_i^z+h\sigma_i^x]}\left\{\sum_{\vec{\tau}(\alpha=2)}|\vec{\tau}(\alpha=4)\rangle\langle\vec{\tau}(\alpha=4)|\right\}\\
&\vdots\\
&\times e^{\frac{\beta}{N_s}\sum_{i=1}^{N}[\lambda(3)\sigma_i^z+h\sigma_i^x]}|\vec{\tau}(\alpha=1)\rangle,\end{aligned}$$

where the last Trotter step is fixed $|\vec{\tau}(\alpha=1)\rangle = |\vec{\tau}(\alpha=1)\rangle$ to fulfill the periodic boundary conditions imposed by the trace. Each of the identities (terms in between {}) can be canceled to obtain

$$\sum_{\vec{\tau}(1),\ldots,\vec{\tau}(N_s)} \prod_{\alpha=1}^{N_s} \langle\vec{\tau}(\alpha)| e^{\frac{\beta}{N_s}\sum_{i=1}^{N}[\lambda(\alpha)\sigma_i^z+h\sigma_i^x]}|\vec{\tau}(\alpha+1)\rangle = \mathrm{Tr}\left[e^{\beta\sum_{i=1}^{N}[(\sum_{\alpha=1}^{N_s}\frac{\lambda(\alpha)}{N_s})\sigma_i^z+h\sigma_i^x]}\right] = N\cos\left(\beta\sqrt{\left(\sum_{\alpha=1}^{N_s}\frac{\lambda(\alpha)}{N_s}\right)^2+h^2}\right). \tag{C4}$$

Going back to Eq. (C1), the factor $N$ in front of the whole expression within the exponential imposes that the exact solution of the partition function is purely dominated by saddle point solution in the large $N$ limit. The saddle point is then determined in terms of all different $m_z(\alpha)$, $\lambda(\alpha)$, one for each Trotter slice $\alpha$. For convenience, let us now denote the magnetization and auxiliary fields as $m_z(\alpha=i) = m_z^i$ and $\lambda(\alpha=i) = \lambda^i$, in order to keep track of the Trotter indices. We then derive each saddle point condition that minimizes the exponential term

$$m_z^i = \frac{\left(\sum_i \frac{\lambda^i}{N_s}\right)}{\sqrt{\left(\sum_i \frac{\lambda^i}{N_s}\right)^2+h^2}}\tanh\left(\beta\sqrt{\left(\sum_i \frac{\lambda^i}{N_s}\right)^2+h^2}\right), \quad \lambda^i = 2Jm_z^i + 4K{m_z^i}^3. \tag{C5}$$

Looking at the first equation for $m_z^i$, we can see how there is no dependence on the Trotter index $i$ on the right-hand side of the expression. This allows us to conclude that the magnetization $m_z^i$ is constant along the imaginary time ($m_z^i \equiv m_z$). The second equation, together with $m_z^i \equiv m_z$, extends this argument to $\lambda^i = \lambda$.

We would like to remark that, when doing such approximation, there is no mixing among Trotter slices. This is exactly what happens when computing a classical partition function, where there are no transverse components. The computation then shows how the transverse fields only enter in a mean-field-like manner, a result expected for long-range interacting systems, which are known to belong to the mean-field universality class [8].